\documentclass[12pt]{article}
\usepackage{graphicx}

\advance\textheight by \topskip
\input epsf

\input{tcilatex}

\begin{document}

\title{\textbf{CMB, Quantum Fluctuations and the Predictive Power of
Inflation}}
\author{ V.F.Mukhanov \\
%EndAName
Sektion Physik, LMU, Theresienstr.37,\\
Munich, Germany e-mail: mukhanov@theorie.physik.uni-muenchen.de}
\maketitle

\begin{abstract}
I comment on the predictive power of inflation and recent CMB measurements.
\end{abstract}

\bigskip

When more than twenty years ago G. Chibisov and I were fortunate to discover
that quantum fluctuations could be responsible for the large scale structure
of the universe, we hardly thought it would be possible one day to verify
this prediction experimentally. We wrote a paper \cite{MC}, published in
JETP Lett, Vol. 33, No.10, p. 532, 20 May 1981\textbf{\ }(see Appendix),%
\textbf{\ }where we derived the spectrum of cosmological metric
perturbations generated in a de Sitter stage of accelerated expansion (the
word "inflation" had not been invented at this time) from quantum
fluctuations. The spectrum came out to be \textit{logarithmically} dependent
on the scale. Our results were obtained for the first particular working
model of inflation based on $R^{2}-$gravity, which is conformally equivalent
to a model with a scalar field\footnote{%
For this reason the calculations in this model are very similar to the
calculations in scalar field models.}.

More than a year later I was very happy to learn that, after some debates,
four groups of people, working on perturbations in the other \textit{%
particular} scenario of inflation, \textquotedblleft new inflation",
estimated the perturbations \cite{H},\cite{S},\cite{GP},\cite{BST} and came
to the same conclusion confirming our earlier result \cite{MC}.

Nowadays, even though there is no reliable particle theory beyond the
standard model, there are many successful inflationary scenarios, most of
which are based on the idea of \cite{L}. The general theory of the
generation of inflationary perturbations which allows us to calculate the
perturbations in these scenarios, where the Hubble constant can change by
orders of magnitude during inflation, was developed in \cite{M1},\cite{M2}
and its current status is described in the review paper \cite{MFB}, which,
to a large extent, is based on \cite{M1},\cite{M2}.

The main lesson we learned from the general theory of inflationary
perturbations is that despite a multitude of various inflationary scenarios,
the spectra of perturbations in the simple models of inflation, that is,
models with minimal number of free parameters, are very similar (namely,
logarithmically dependent on scale), and in complete agreement with earlier
calculations \cite{MC}. Contrary to an erroneous belief inflation does not
predict the scale-invariant, Harrison-Zel'dovich spectrum. The spectral
index should be in the range of $0.92<n_{s}<0.97.$ The physical reason for
the deviation of the spectrum from the scale invariance is the necessity of
having a graceful exit from inflation.

Can one avoid this\textit{\ prediction} of inflation? In principle yes, but
only by spoiling \textit{the most predictive model, namely, simple
inflationary model }(not be be confused with any one particular scenario).
To avoid the main prediction of inflation, i.e. a logarithmically dependent
spectrum, one has to introduce unjustified extra parameters and perform some
fine tuning. Of course, this diminishes to a large extent the predictivity
of inflation and, taken to extremes, one "can explain" nearly any outcome of
any measurement in this way.

The recent precise measurements of the CMB fluctuations, especially by WMAP,
seem to be in a very good agreement with the predictions of simple
inflationary theory \cite{WMAP}. It is still too early to talk about a
reliable detection of logarithmic deviations of the spectrum from the flat
one because the measurements are not yet sufficiently accurate. However, the
progress in this direction is very impressive and future experiments should
be able to verify this very \textit{robust prediction} of inflation.

Does inflation have any competitor in explaining the outcome of the CMB
measurements? Some authors claim that the cyclic universe scenario predicts
the same spectrum as inflation. I believe that comparison of these models is
misleading. The cyclic model does not predict \textit{anything }for an
expanding universe. In this scenario the spectrum comes to be nearly flat
only at the last stage of collapse of the universe, before the singularity.
The main obstacle in transferring the spectrum to an expanding branch is the
singularity, which makes most dubious not only the conclusions about
generated perturbations, but also the whole cyclic scenario. At present,
there is not slightest hint how the singularity problem, which seems much
more difficult than all other cosmological problems, could be solved.
Therefore statements like \textquotedblleft the cyclic universe predicts the
perturbations spectrum" are completely unjustified and can be considered
only as a wishful thinking.

Returning to inflationary cosmology, it is still too early to say that
simple inflation has been proved by observations. However, it seems that we
are on the right track to confirm the most dramatic prediction of
inflationary cosmology that the structure of the universe and hence our
lives are due to quantum fluctuations.

The first paper where the spectrum of inflationary perturbations was
calculated \cite{MC} is not readily available. In order to make this paper
easily accessible to those who want to study the inflationary cosmology, I
attach it below.

\newpage

\appendix

\bigskip

\begin{center}
\textbf{JETP Lett, Vol. 33, No.10, 20 May 1981}
\end{center}

\section*{Quantum fluctuations and a nonsingular Universe}

\begin{center}
V.F.Mukhanov and G.V. Chibisov

P. N. Lebedev Physics Institute, Academy of sciences of the USSR

(Submitted 26 February 1981; 15 April 1981)

\bigskip

Pis'ma Zh. Eksp. Theor. Fiz. 33, No.10, 549-553 (20 May 1981)
\end{center}

\bigskip

\textbf{Abstract}. Over a finite time, quantum fluctuations of the curvature
disrupt the nonsingular cosmological solution corresponding to a universe
with a polarized vacuum. If this solution held as an intermediate stage in
the evolution of the universe, then the spectrum of produced fluctuations
could have lead to formation of galaxies and galactic clusters.

\bigskip

PACS numbers: 98.80.Bp,9850.Eb

\bigskip

\ \ A nonsingular cosmological model with a polarized vacuum has been
attracting particular interest recently \cite{Star}. It has been pointed out
elsewhere that quantum fluctuations may prove important in cosmology at
energy density comparable to the Planck value \cite{Ginz}. Since this are in
fact the energy densities characteristic of the nonsingular polarized-vacuum
model \cite{Star}, we believe it is worthwhile to study the role of quantum
fluctuations in order to determine whether there is a singularity in this
model.

For an isotropic metric, the single-loop corrections describing the
polarization of vacuum of physical fields in a strong gravitational field
lead to the following Einstein equations \cite{BD}:%
\begin{eqnarray}
R_{k}^{i}-\frac{1}{2}\delta _{k}^{i}R &=&\frac{1}{H^{2}}\left(
R_{l}^{i}R_{k}^{l}-\frac{2}{3}RR_{k}^{i}-\frac{1}{2}\delta
_{k}^{i}R_{m}^{l}R_{l}^{m}+\frac{1}{4}\delta _{k}^{i}R^{2}\right) -
\label{1} \\
&&-\frac{1}{6M^{2}}\left( 2R_{\text{ \ };k}^{;i}-2\delta _{k}^{i}R_{\text{ \ 
};l}^{;l}-2RR_{k}^{i}+\frac{1}{2}\delta _{k}^{i}R^{2}\right) ,  \nonumber
\end{eqnarray}%
where the coefficients $M^{2}$ and $H^{2}$ result from the sum over the
effects of all the fields. For stability of Minkowski space, $M^{2}$ must be
positive. For $H^{2}>0,$ Eqs. (\ref{1}) have a particular solution of the de
Sitter type \cite{Star},%
\begin{equation}
ds^{2}=g_{ik}dx^{i}dx^{k}=a^{2}\left( \eta \right) \left( d\eta
^{2}-\dsum\limits_{\alpha =1}^{3}\left( dx^{\alpha }\right) ^{2}\right) ,
\label{2}
\end{equation}

\[
a\left( \eta \right) =-\frac{1}{H\eta },\text{ \ }-\infty <\eta <0,\text{ }%
R=-12H^{2}=const. 
\]%
The curvature invariants, in particular $R,$ have no singularities in the
limit $\eta \rightarrow -\infty ,$ showing that there is no actual
singularity in the universe described by metric (\ref{2}).

Let us consider, against the background of this metric, some small
perturbations that satisfy Eqs. (\ref{1}). We restrict the discussion to
scalar perturbations. In a synchronous frame of reference $\left( \delta
g_{\alpha 0}=0\right) $ these perturbations have the following tensor
structure:%
\begin{equation}
h_{\beta }^{\alpha }=-\frac{1}{a^{2}}\delta g_{\alpha \beta }=\left( \nabla
^{\alpha }\nabla _{\beta }-\frac{1}{3}\delta _{\beta }^{\alpha }\Delta
\right) \lambda -\frac{1}{3}\delta _{\beta }^{\alpha }\Delta \mu ,  \label{3}
\end{equation}%
where $\nabla ^{\alpha }=\nabla _{\alpha }=\partial /\partial x^{\alpha },$
and $\Delta $ is the Laplacian.

The convolution of Eqs. (\ref{1}), linearized with respect to $h_{\beta
}^{\alpha },$ yields a second order equation for perturbations of the
curvature scalar: $\delta R:$%
\begin{equation}
\delta R^{\prime \prime }+2\frac{a^{\prime }}{a}\delta R^{\prime }-\Delta
\delta R-M^{2}a^{2}\delta R=0,  \label{4}
\end{equation}%
where the prime denotes differentiation with respect to $\eta ,$ and $%
a\left( \eta \right) =-1/H\eta .$

Also using Eqs. (\ref{1}), we can show that all the quantities in which we
are interested (in particular, the $h_{\beta }^{\alpha }$), can be expressed
in terms of $\delta R.$ This quantity plays a special role because of its
invariance with respect to so called fictitious perturbations \cite{Lif}
which stem from transformations of the coordinate system which does not
disrupt the synchronism. Fictitious perturbations of scalar functions result
from the change in the origin for the time scale \cite{zel}; since $R$ is
independent of the time ($R=const$) in the de Sitter model, with which we
are concerned here, we have $\delta R=0$ for these fictitious modes.

Adopting a perturbation of the curvature scalar as a physical variable, we
find the corresponding action in the form\cite{mu}%
\begin{equation}
\delta S_{b}=\frac{1}{2}\int d^{4}x\left[ \phi ^{\prime 2}-\nabla ^{\alpha
}\phi \nabla _{\alpha }\phi +\left( \frac{a^{\prime \prime }}{a}%
+M^{2}a^{2}\right) \phi ^{2}\right] ,  \label{5}
\end{equation}%
where $\phi =1/\sqrt{18\left( 4H^{2}-M^{2}\right) }$ $a\delta R/M\ell ,$ and 
$\ell =\left( 8\pi G/3\right) ^{1/2}=4.37\times 10^{-33}$ $cm$ is the Planck
length.

In quantizing we should note that the physical system under consideration (a
polarized vacuum in a gravitational field) has a finite energy density,
which must undergo fluctuations according to the uncertainty principle.
These fluctuations are zero-point oscillations of the field of collective
excitations of the ordinary physical fields (``scalarons'' with mass $M).$

The canonical quantization is carried out by a procedure similar to that of
Ref. \cite{Grib}. As a result we can calculate various correlation
functions, e.g., 
\begin{equation}
\left\langle 0\left\vert \delta \hat{R}\left( \mathbf{x}\right) \delta \hat{R%
}\left( \mathbf{x+r}\right) \right\vert 0\right\rangle =\frac{1}{2\pi ^{2}}%
\int P^{2}\left( k\right) \frac{\sin kr}{kr}\frac{dk}{k},  \label{6}
\end{equation}%
where $k$ and $r$ are the physical wave vector and the physical length,
expressed in units of reciprocal centimeters and centimeters, respectively.
The spectrum $P\left( k\right) $ is expressed in terms of Bessel functions
of index $3\nu /2=\left( 3/2\right) \sqrt{1+\left( 4/9\right) M^{2}/H^{2}}$
and argument $k/H,$ which are the exact solutions of Eq. (\ref{4}). The
spectrum $P\left( k\right) $ is shown in Fig.1. A distinguishing feature of
this spectrum is a maximum at $k\sim H\left( \eta /\eta _{0}\right) $ ($\eta
_{0}$ is the initial time at which the vacuum of the scalaron field is
given); the perturbation amplitude at this maximum increases over time. Over
a finite time, which in the most interesting case ($M^{2}\ll H^{2}$) is%
\begin{equation}
\delta t_{f}=\frac{3}{2H}\frac{H^{2}}{M^{2}}\ln \left( \frac{1}{2\left( \ell
M\right) ^{2}}\right)  \label{7}
\end{equation}%
the amplitude of the curvature perturbations at the maximum reaches
characteristic of the original background model, and the universe enters a
stage of Friedmann expansion with ordinary hydrodynamic matter as a result
of multiple production of scalarons with finite wave numbers (primarily with 
$k=$ $H\left( \eta /\eta _{0}\right) $). Thus we see that the quantum
fluctuations, which are necessary present in the system, cause the universe
to spend a finite time in the de Sitter state and thus cast doubt on the
possibility of a nonsingular beginning of the universe. Regardless of the
nature of singularity (classical or quantum), we believe that this fact
significantly detracts from the esthetic value of the original model.

\begin{figure}[h!]
\centering\leavevmode\epsfysize=10cm \epsfbox{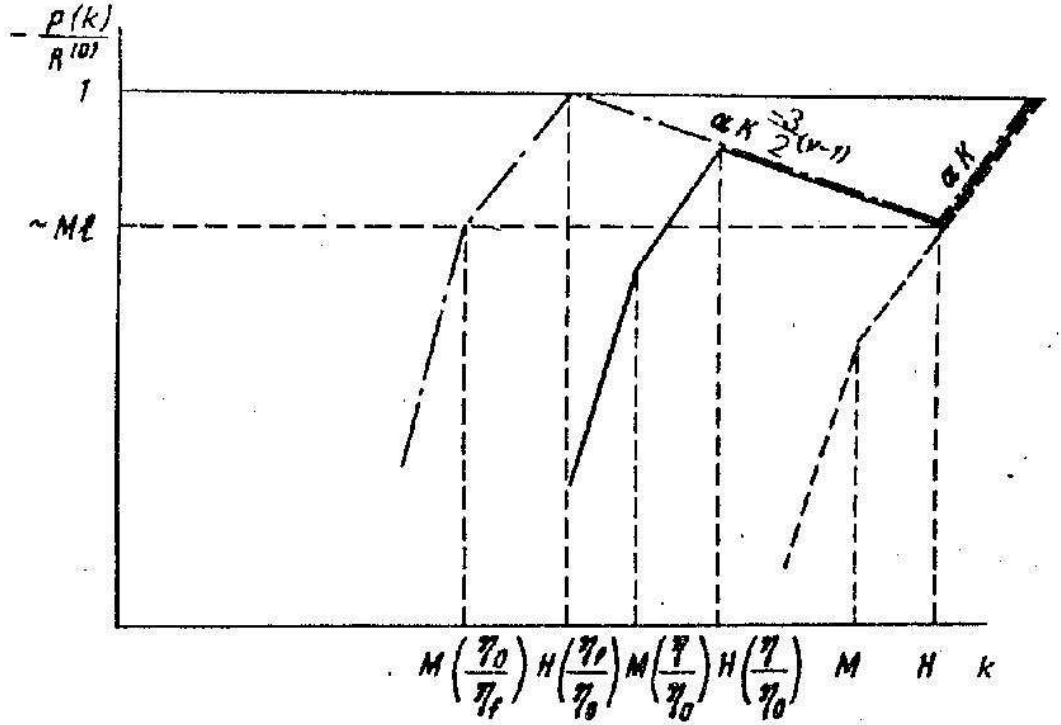}
\caption[fig1]{Spectrum of relative curvature fluctuations, $P(k)/R^{(0)}$,
plotted as a function of the wave number $k$. Dashed lines - spectrum of
vacuum fluctuations which is specified at some initial time $\protect\eta_0$%
; solid line - the spectrum into which the vacuum spectrum transforms at
some later time. }
\label{fig1}
\end{figure}

A finite duration of the de Sitter stage does not by itself rule out the
possibility that this stage may exist as an intermediate stage in the
evolution of the universe. An interesting question arises here: Might not
perturbations of the metric , which would be sufficient for the formation of
galaxies and galactic clusters, arise in this stage? To answer this
question, we need to calculate the correlation function for the fluctuations
of the metric after the universe goes from the de Sitter stage to the
hydrodynamic stage. By analogy with (\ref{6}) we find 
\begin{equation}
\left\langle 0\left\vert \hat{h}\left( \mathbf{x}\right) \hat{h}\left( 
\mathbf{x+r}\right) \right\vert 0\right\rangle =\frac{1}{2\pi ^{2}}\int
Q^{2}\left( k\right) \frac{\sin kr}{kr}\frac{dk}{k},  \label{8}
\end{equation}%
where $h=h_{\alpha }^{\alpha }$ and where, for the most interesting region, $%
H>k>H\exp \left( -3H^{2}/M^{2}\right) $ $\left( M^{2}\ll H^{2}\right) ,$%
\begin{equation}
Q\left( k\right) \approx 3\ell M\left( 1+\frac{1}{2}\ln \frac{H}{k}\right) .
\label{9}
\end{equation}

The fluctuation spectrum is thus nearly flat. The quantity $Q\left( k\right) 
$ is the measure of the amplitude of perturbations with scale dimensions $%
1/k $ at the time the universe begins the ordinary Friedmann expansion. With 
$\ell M\sim 10^{-3}-10^{-5}$ and $M/H\leq 0.1-$these values are consistent
with modern theories of elementary particles-the amplitude of the
perturbations of the metric on the scale of galactic clusters turns out to
be equal to $10^{-3}-10^{-5},$ and these perturbations can lead to the
observed large scale structure of the universe. The form of spectrum (\ref{9}%
) is completely consistent with modern theories for the formation of
galaxies \cite{zel}.

To summarize: Using a de Sitter model as an example, we have shown that
quantum fluctuations (zero-point vibrations) cause the universe to spend a
finite time in a state with polarized vacuum. This result casts doubt on the
possibility of a nonsingular origin for the universe. However, models in
which the de Sitter stage exists only as an intermediate stage in the
evolution are attractive because fluctuations of the metric sufficient for
the galaxy formation can occur. Thus we have one possible approach for
solving the problem of the appearance of the original perturbation spectrum.

We thank V. L. Ginzburg, Ya. B. Zel'dovich, M. A. Markov, and A. A.
Starobinskii for discussions.

{\small Translated by Dave Parsons}

{\small Edited by S. J. Amoretty}

\end{document}